# Single-Switch-Regulated Resonant WPT Receiver

Kerui Li, *Student Member, IEEE*, Siew-Chong Tan, *Senior Member, IEEE,* and Ron Shu Yuen Hui, *Fellow, IEEE*

*Abstract*-A single-switch-regulated wireless power transfer (WPT) receiver is presented in this letter. Aiming at low-cost applications, the system involves only a single-switch class-E resonant rectifier, a frequency synchronization circuit, and a microcontroller. The number of power semiconductor devices required in this circuit is minimal. Only one active switch is used and no diode is required. As a single-switch solution, this simplifies circuit implementation, improves reliability, and lowers hardware cost. The single-switch resonant rectifier provides a relatively constant quasi-sinusoidal voltage waveform to pick up the wireless power from the receiver coil. Due to the resonant nature of the rectifier, ZVS turn on and turn off are achieved. The steady-state analysis and discussions on the component sizing and the control design are provided. A prototype is built and experimental works are performed to verify the features: achieving 93% efficiency single-switch AC-DC rectification (85% including the auxiliary circuit), −7.17dB THD AC voltage waveform, ZVS turn on and turn off over wide load range, 0.4% output voltage regulation error, 79% cost reduction as compared with the Qi- compliant receivers, 2.5% overshoot/undershoot after load disturbances, and 1% overshoot after line disturbance.

## I. INTRODUCTION

The consumer electronics industry is experiencing rapid development over the last few years. Many portable electronic devices are shifting from cable charging to wireless charging. Several works on cost-effective and reliable wireless power transfer (WPT) receiver have been reported [1], [2]. In the context of cost-effective and reliable power electronics design, the number of power semiconductor devices is one of the critical concerns. Consequently, minimizing the number of power semiconductor devices in designing a regulated WPT receiver has been explored and is presented in this letter.

In terms of high-frequency rectification, the passive diode bridge rectifier with a post regulator is the most common solution [3]. While generally straightforward and effective, this is a relatively expensive and unreliable solution as it involves quite many power semiconductor devices. Moreover, due to the discontinuous conduction mode (DCM) operation of the diode bridge at light load power, considerable current surge is generated at the transmitter coil of the WPT system [4]. This may lead to overcurrent of the transmitter switches and overvoltage of its compensated capacitor, thereby aggravating the reliability issue.

The reconfigurable diode-bridge rectifier (with five power switches) has been proposed as a high-frequency WPT receiver to replace the use of full-bridge diode rectifier [5] in consumer electronic applications. The DC regulation is achieved by mode switching at particular moments, which leads to discontinuous output regulation and extra low-frequency output ripples. To further reduce the use of active power switches, the active full-bridge rectifier (with four power switches) has been proposed [6]–[8]. It can achieve high-frequency DC regulation without a post regulator and can avoid the current surge issue. However, the high switching frequency operation imposes demanding challenges on the synchronization of the high-side and low-side switches, of which failure will lead to a direct short circuit of the source or load. For safe operation, introducing dead time into the switching process is inevitable. Nevertheless, this is not a fool-proof solution and possible electrical faults of the switch or driver, as well as signal noise, may still cause a direct short circuit of a bridge circuit that operates at high frequency.

Alternatively, a simpler active rectifier solution based on the half-bridge rectifier can be used [9]. It provides a similar function to that of the full-bridge rectifier, but possesses a simpler structure and lower cost. Nevertheless, comprising still with a half-bridge inverter circuit, it suffers from the same reliability issues as those of the active full-bridge rectifier. To resolve this, it is necessary to adopt topologies that do not contain the half-bridge inverter structure. For this reason, a single-switch solution would be preferred in terms of achieving low cost and improved reliability.

Notably, the active class E rectifier is currently the only possible candidate for achieving a single-switch rectifier solution. Unfortunately, the use of class E rectifier to achieve DC regulation is rarely reported. Incorporation of a post regulator is the common approach towards addressing the regulation issue [10]-[12]. However, this again is cost ineffective. It, therefore, raises the question as to whether it is possible to realize high-frequency rectification and regulation altogether using a truly single-switch solution.

In this letter, our investigation on the feasibility of using a single-switch solution based on the class E rectifier in realizing simultaneously the high-frequency rectification and DC regulation, is reported. Our proposed WPT receiver system comprises only a single-switch class-E resonant rectifier, a frequency synchronization circuit, and a microcontroller. It uses the minimum number of power semiconductor devices possible (i.e., one power MOSFET) and no diode is required. The frequency synchronization circuit is used to lock up the frequency and phase of the current of the receiver coil. By using the microcontroller to control the phase-shift angle between the AC voltage of the rectifier and the current of the receiver coil, the real input power can be controlled, which thereby achieves the regulation of the DC output voltage. To achieve such control, a new design method is proposed and will be discussed in this letter.

## II. PROPOSED SINGLE-SWITCH REGULATED RESONANT WPT RECEIVER SYSTEM

### A. Circuit Configuration

As depicted in Fig. 1, the single-switch regulated resonant WPT receiver involves the compensated receiver coil ($L_s$ and $C_s$), the single-switch resonant rectifier circuit ($C_f$, $L_f$, $C_o$, $R$ and $S$), the synchronization circuit, and the microcontroller (MCU).

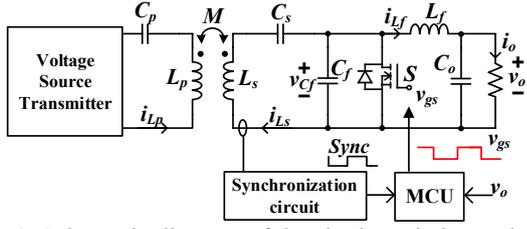

Fig. 1. Schematic diagram of the single-switch-regulated resonant WPT receiver system.

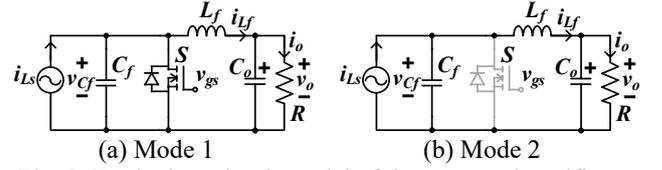

(a) Mode 1    (b) Mode 2

Fig. 3. Equivalent circuit model of the proposed rectifier.

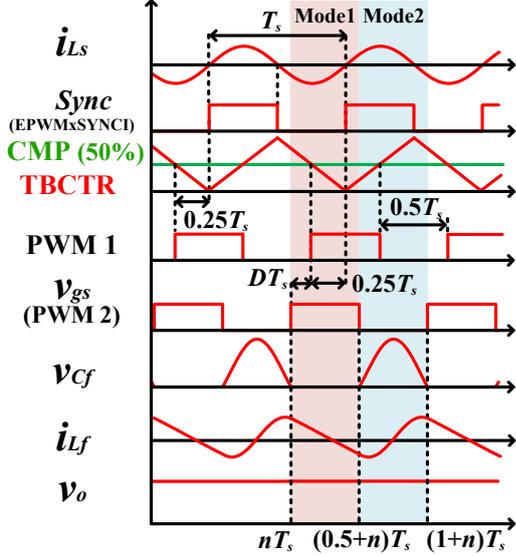

Fig. 2. Key waveforms of the single-switch-regulated

As compared to conventional class E rectifiers [10], the inductance $L_f$ of the proposed system is significantly reduced, enabling the use of air core inductor. Moreover, the voltage waveform of the main switch ($v_{Cfs}$) is invariant to load change. This assures that voltage stress of the main switch is constant and ZVS operation is achievable over a wide load range.

*B. Operating Principles*

In elaborating the operating principles, the following assumptions are adopted:
1) the equivalent series resistance of the reactive components are sufficiently small and negligible;
2) the quality factor of the resonant tank $L_sC_s$ is sufficiently high, of which only the fundamental component of the electrical current can be passed;
3) the output capacitor $C_O$ is relatively large such that its voltage is a relatively constant DC output.

Due to the nature of series-series compensation, the phase angle and amplitude of the secondary current $i_{Ls}$ are only dependent on the transmitter side AC voltage and the coupling coefficient $M$, and not the load [14]. Therefore, its amplitude and phase angle will remain constant throughout the operation. The key waveforms of the single-switch-regulated resonant WPT receiver are depicted in Fig. 2. The synchronization circuit is operated to detect the zero crossing point of the current $i_{Ls}$ and sends a square-pulse synchronization signal *sync* to the microcontroller. This provides the phase angle and frequency information of $i_{Ls}$. With this information, the synchronized PWM signal $v_{gs}$ that has 50% duty cycle and the phase-shift ratio of $D$ ($0 \leq D \leq 0.25$), is generated. There are two operation modes for this circuit: Mode 1 occurs in the time period of $t$ where $nT_s \leq t < (0.5+n)T_s$, and Mode 2 occurs in the time period $t$ where $(0.5+n)T_s \leq t < (1+n)T_s$ for $n = 0, 1, 2, \ldots$. The corresponding equivalent circuit models are shown in Fig. 3.

*Mode 1* [At time $nT_s \leq t < (0.5+n)T_s$]

In Mode 1 (see Fig. 3(a)), the main switch $S$ is in the on-state. The current $i_{Ls}$ of the wireless coil is freewheeled through switch $S$. Consequently, the voltage of capacitor $C_f$ remains zero, i.e.,

$$v_{Cf} = 0 \quad (1)$$

Meanwhile, inductor $L_f$ discharges its stored energy linearly with a current of $i_{Lf}$ to the output capacitor $C_o$, of which provides a relatively constant current to load $R$, such that

$$L_f \frac{\partial i_{Lf}}{\partial t} = -v_o \quad (2)$$

$$C_o \frac{\partial v_o}{\partial t} = i_{Lf} - \frac{v_o}{R} \quad (3)$$

At time $t=(0.5+n)T_s$, switch $S$ is turned off under ZVS turn off as $v_{cf}$ is zero volt.

*Mode 2* [At time $(0.5+n)T_s \leq t < (1+n)T_s$]

In Mode 2 (see Fig. 3(b)), switch $S$ is in the off-state. Here, current $i_{Ls}$ flows through the resonant tank $L_f C_f$, such that

$$C_f \frac{\partial v_{Cf}}{\partial t} = i_{Ls} - i_{Lf} = -|I_{Ls}|\cos(2\pi f_s t - 2\pi D) - i_{Lf} \quad (4)$$

$$L_f \frac{\partial i_{Lf}}{\partial t} = v_{Cf} - v_o \quad (5)$$

The resonant frequency of the resonant tank $L_f C_f$ is very close to the switching frequency ($f_s$). Thus, the waveform of $v_{Cf}$ is sinusoidal-like, involving the resonance of $L_f$ and $C_f$. With the gradual increase of voltage on $C_f$, the ZVS turn-on condition is achieved. Meanwhile, current $i_{Lf}$ concurrently charges the output capacitor $C_o$ and provides current to load $R$, such that

$$C_o \frac{\partial v_o}{\partial t} = i_{Lf} - \frac{v_o}{R} \quad (6)$$

At time $t=nT_s$, the voltage level of $v_{cf}$ reaches zero (after resonance). The switch is turned on at this instance, thereby achieving ZVS turn on.

By solving the equations above, $v_{Cf}$ and $i_{Lf}$ can be derived as

$$v_{Cf} \approx \begin{cases} 0 & nT_s \leq t \leq (n+0.5)T_s \\ 2.26v_o \cos(8.11/T_s (t - 0.75T_s)) + v_o & (n+0.5)T_s \leq t \leq (n+1)T_s \end{cases} \quad (7)$$

$$i_{Lf} \approx \begin{cases} 0.795|I_{Ls}|\sin(2\pi D) - 8.11v_o \sqrt{C_f/L_f}(t/T_s - 0.25) & nT_s \leq t \leq (n+0.5)T_s \\ 0.795|I_{Ls}|\sin(2\pi D) + 2.26v_o \sqrt{C_f/L_f}\sin(8.11/T_s(t - 0.75T_s)) & (n+0.5)T_s \leq t \leq (n+1)T_s \end{cases} \quad (8)$$

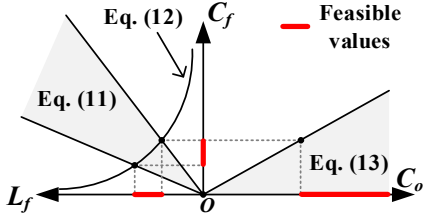

Fig. 4. Feasible region of the parameter design.

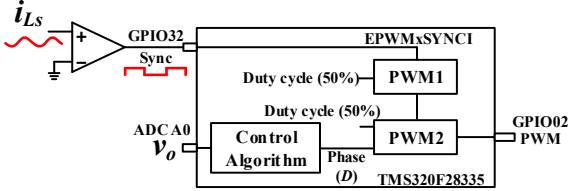

Fig. 5. Frequency and phase angle synchronization.

The real power $P$ received by the receiver coil is

$$P = \frac{1}{T_s} \int_0^{T_s} v_{Cf} i_{Ls}\, dt \approx 0.795 |I_{Ls}| v_o \sin(2\pi D) \quad (9)$$

By applying charge balance on capacitor $C_O$ (see (6)), the output current and voltage can be obtained as

$$\begin{cases} i_o = 0.795 |I_{Ls}| \sin(2\pi D) \\ v_o = 0.795 |I_{Ls}| R \sin(2\pi D) \end{cases} \quad (10)$$

Eq. (10) indicates that the output current $i_o$ or the output voltage $v_o$ can be regulated by adjusting $D$. Note that the output current is load independent. From the perspective of the load, the output from the rectifier is a constant current source, which is good for constant current charging. In contrast, because of this current source nature, the output voltage is highly load-dependent, and a feedback control loop is required to regulate the output voltage.

## III. DESIGN CONSIDERATIONS

### A. Design of Resonant Rectifier ($L_f$, $C_f$, and $C_o$)

Two design objectives are imposed when designing $L_f$ and $C_f$: 1) AC voltage waveform of $v_{Cf}$ must be maintained relatively constant for all loads; 2) ZVS operation must be enforced for all loads.

As shown in (4), the voltage waveform $v_{Cf}$ involves the contribution of the resonant current $i_{Lf}$, load current ($v_o/R$), and the current of the receiver coil $i_{Ls}$. To ensure a relatively constant waveform regardless of the load and the current from the receiver coil (1st design objective), the characteristic admittance $\sqrt{C_f/L_f}$ has to be sufficiently large so that the resonant current (highlighted in (8)) is dominant. However, the increased characteristic admittance will lead to large reactive power and unnecessary conduction loss. As a result, the characteristic admittance is designed as

$$5 \frac{\max|I_{Ls}|}{\min\{v_o\}} \geq \sqrt{C_f/L_f} \geq 2.5 \frac{\max|I_{Ls}|}{\min v_o} \quad (11)$$

where $\max(|I_{Ls}|)$ and $\min(v_o)$ are the maximum and minimum values of the current of receiver coil $i_{Ls}$ and output voltage $v_o$. To achieve ZVS operation, the voltage waveform of $v_{Cf}$ should be zero at time $0.5T_s$ and $T_s$, i.e. $v_{Cf}(0.5T_s) = v_{Cf}(T_s) = 0$. Meeting ZVS operation requirement (2nd design objective), the resonance frequency of $L_f$ and $C_f$ is then designed as

$$\frac{1}{2\pi \sqrt{L_f C_f}} = 1.29 f_s \quad (12)$$

Using (11) and (12), the corresponding values of $L_f$ and $C_f$ are obtained.

The design objective of $C_o$ is to reduce the output ripple to be less than $x\%$ of the nominal value. According to (6), the output ripple is contributed by inductor current $i_{Lf}$ and output current ($v_o/R$). Consequently, the output capacitor is designed as

$$C_o \geq \frac{5.41 C_f}{x\%} \quad (13)$$

As shown in Fig. 4, a set of feasible values $L_f$ and $C_f$ can be obtained by solving the cross section points of (11) and (12). Then, the minimum value of $C_o$ is calculated via (13) using the maximum value of $C_f$.

### B. Frequency and Phase Synchronization

As shown in Fig. 5, the external circuit sends the synchronization signal *sync* to MCU TMS320F28335 via GPIO32, which is specified for PWM synchronization. After receiving the synchronization signal, the frequency of the PWM modules is synchronized. As shown in Fig. 2, the time-based counter TBCTR restarts counting from zero after detecting the rising edge of the synchronization signal *sync*. Using the counter-compare (CMP) and TBCTR, synchronized PWM 1 with 50% duty cycle is generated. The phase angle difference between PWM 1 and the resonant current $i_{Ls}$ is a quarter of a switching cycle ($=0.25 T_s$). Taking PWM 1 as the reference, PWM 2 utilizes phase-shift ratio $D$, which is computed by the control algorithm, to generate the required phase-shift PWM signal $v_{gs}$ (PWM 2). As a result, the total phase shift between $v_{gs}$ and $i_{Ls}$ is $(0.25+D)T_s$. The resulted phase-shift PWM is then used to drive the circuit into regulating the DC output voltage.

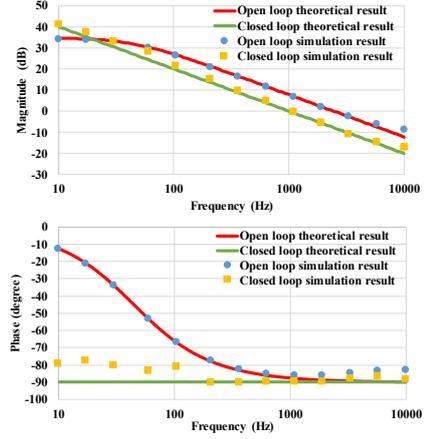

Fig. 6. Bode plots of the receiver system.

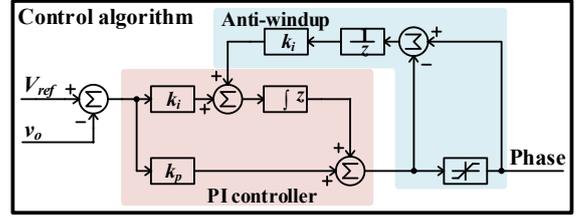

Fig. 7. Block diagram of the control algorithm.

TABLE I. A COMPARISON WITH EXISTING WPT RECEIVERS

| | **Proposed** | Qi compliant receiver [3] | [6] | [7] | [12] |
|---|---|---|---|---|---|
| Topology | **Single-switch resonant rectifier** | Diode rectifier & Linear regulator | Reconfigurable diode bridge rectifier | Active full bridge rectifier | Passive Class E rectifier & DC-DC converter |
| Compensation of receiver coil | **Series compensation** | Series parallel compensation | Series compensation | Series compensation | Series parallel compensation |
| Resonant frequency | **200 kHz** | 100 kHz | 6.78 MHz | 35 kHz | 13.56 MHz |
| Number of power switches (including diodes) | **1** | 5 | 4 | 4 | 3 |
| Maximum Voltage stress | **≈3$V_o$** | $V_{DC\text{-}link}(>V_o)$ | $V_o$ | $V_o$ | $V_{DC\text{-}link}(>V_o)$ |
| Regulation method | **Synchronized Phase-shift control** | Linear regulator | Synchronized switching mode control | Synchronized PWM control | DC-DC converter |
| Design complexity | **High** | Low | High | Medium | Medium |
| Soft switching operation | **ZVS turn on and off** | Diode: ZCS turn on and off; | ZCS turn on and off | Hard switching | ZVS turn on and turn off DC-DC converter: Hard switching |
| Output voltage, power | **24 V, 16 W** | 5.3 V, 5.3 W | 5 V, 6 W | 400 V, 3 kW | 5 V, 13 W |
| Maximum Efficiency | **93 %(Power stage) 85 %(Including auxiliary circuit)** | 76 %(Including auxiliary circuit) | 92 %(Including auxiliary circuit) | 98 % (Power stage) | 87% (Power stage) |
| Direct short circuit risk | **No** | No | Yes | Yes | No |
| Cost of the prototype* | **HKD$342.79** | HKD$1604.44 | N/A | N/A | HKD$374.9 |

*The cost of the prototype is the total price of purchasing discrete components (including power circuit, auxiliary circuit and controller) at unit price, all from the same component distributor.
**The power loss of auxiliary circuit is not optimized.

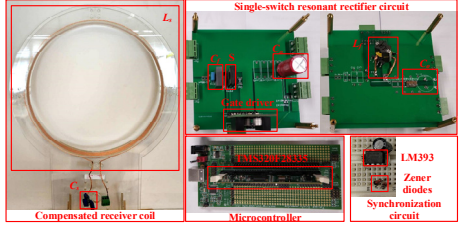

Fig. 8. Photograph of the prototype.

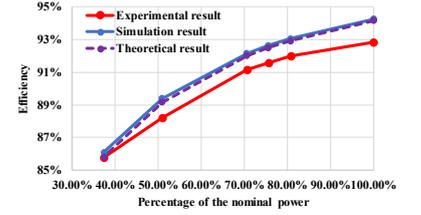

Fig. 9. Efficiency of the receiver.

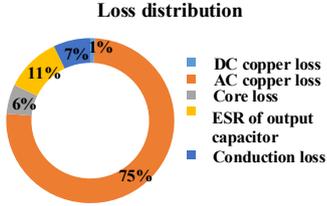

Fig. 10. Loss distribution of the power stage.

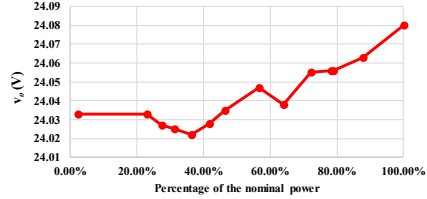

Fig. 11. Output voltage regulation for variable power.

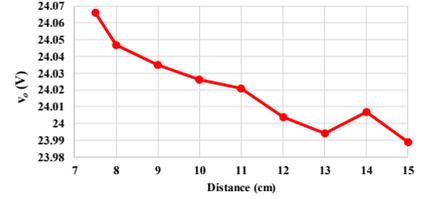

Fig. 12. Output voltage regulation for variable distance.

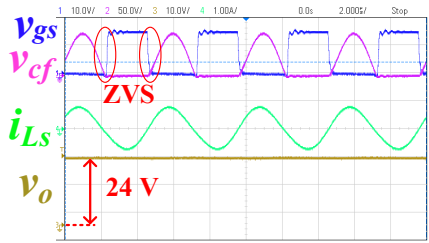

Fig. 13. Steady-state waveforms.

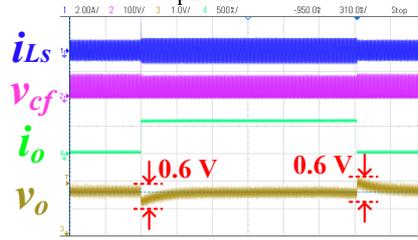

Fig. 14. Dynamic response for $i_o$ steps.

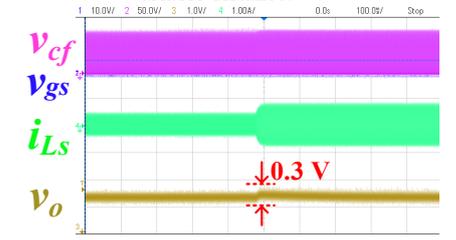

Fig. 14. Dynamic response for $i_{Ls}$ step.

*C. Output Regulation*

By neglecting the high-frequency term in (6), the system equation can be derived as

$$C_O \frac{\partial v_o}{\partial t} = 0.795 |I_{Ls}| \sin(2\pi D) - \frac{v_o}{R} \quad (14)$$

It is observed that the main pole of the receiver system is dominated by the value of $C_o$. By linearizing (14) with consideration to AC perturbation of $D$, the resulting small-signal linearized equation is

$$C_O \frac{\partial \widetilde{v_o}}{\partial t} = 5|I_{Ls}| \cos(2\pi D)\widetilde{D} - \frac{\widetilde{v_o}}{R} \quad (15)$$

The theoretical and the simulation results are very close at the low-frequency range of ≤1000 Hz, thereby validating the accuracy of the derived small-signal equation (15). A proportional-integral (PI) compensator with an anti-windup loop, as shown in Fig. 7, is used as the voltage feedback control loop. The corresponding $k_p$ and $k_i$ of the PI compensator are designed as

$$k_p = \frac{2\pi f_c C_O}{5|I_{Ls}|\cos(2\pi D)}, k_i = \frac{k_p}{RC_O} \quad (16)$$

where $f_c$ is the desired crossover frequency, $R$ is the equivalent resistance of the nominal load, and $D$ is the nominal phase-shift ratio. Therefore, using the open loop system transfer function $G_s(s)$ (15) and the transfer function of the PI compensator $G_{PI}(s)$, the closed-loop transfer function $T(s)$ is obtained as

$$T(s) = G_s(s)G_{PI}(s)$$
$$= \frac{5|I_{Ls}|R\cos(2\pi D)}{1+sRC_O} \frac{2\pi f_c C_O}{5|I_{Ls}|\cos(2\pi D)} \frac{1+sRC_O}{sRC_O} = \frac{2\pi f_c}{s} \quad (17)$$

The Bode plots of the open loop transfer function (15), the compensated transfer function $T(s)$ of the system (17), and the numerical results obtained from circuit simulation are shown in Fig. 6.

## IV. COMPARISON

A comparison of the proposed system with existing WPT receivers is provided in Table I. These solutions are compared in terms of topologies, compensation of receiver coil, resonant frequency, etc. As illustrated, the proposed receiver is the only single-switch solution with output regulation. Its inherent ZVS turn on and off operations make it an excellent candidate for MHz WPT applications. The elimination of the post regulator further simplifies the architecture of the receiver system. In addition, with the single-switch configuration, the cost of the proposed solution is reduced and the related direct short circuit risk is prevented. The advantages over the two-stage solutions [3], [11], are that the proposed solution achieves both regulation and AC-DC rectification in one stage with the minimum number of power switches. The advantages over the single-stage full-bridge based solution [6] and [7], are that the proposed solution can realize high-frequency DC regulation and secure soft switching.

## V. EXPERIMENTAL RESULTS

A 200 kHz switching frequency, 24 V output, 16 W prototype, as shown in Fig. 8, is built, with parameters $C_f$ =76 nF, $L_f$=5.3 $\mu$H, $L_s$=164 $\mu$H, $C_r$=3.86 nF, and $C_o$=3300 μF. Fig. 9 illustrates the calculated, simulated and measured efficiency curves of the receiver. The maximum achievable efficiency in the experiment is 93%. Fig. 10 shows a chart of the loss distribution of the power stage. Fig. 11 and 12 show respectively the curves of the output voltage versus the output power, and that of the output voltage versus the distance between two coils. The output voltage error for different operating conditions is always less than 0.1 V (0.4% of the nominal voltage), which verifies that the receiver is well regulated in either variable load or variable coupling operation. Fig. 13 shows the steady-state waveform of the receiver. It is observed that both ZVS on and off operations are achieved. Fig. 14 shows the dynamic response for load step changes. The output voltage has an overshoot of 0.6 V (2.5% of nominal voltage) after the load power steps from 0 to full power. A undershoot of 0.6 V (2.5% of nominal voltage) is present after the load power steps down from full power to 0. Fig. 15 shows the dynamic response of the system with the peak-to-peak value of $i_{Ls}$ stepping from 1 A to 1.6 A. A 0.3 V overshoot (≈1% of nominal $v_s$) is observed before it is regulated back to zero error. These results verify that not only is the output voltage well regulated in steady state at its nominal value with a relatively low error, but it is also dynamically stable and well regulated against disturbances of load power and input current changes.

## VI. CONCLUSIONS

This letter reports and shows that it is feasible to achieve a single-switch regulated resonant WPT receiver using the active class E rectifier topology. Our prototype and results validate that by adopting phase-shift control and our proposed design methodology, the WPT system can simultaneously achieve high-frequency rectification, wide-load-range ZVS operation, and accurate output regulation, without using any additional post-regulator. As the system involves only one power semiconductor device (no diode required), lower cost and higher reliability as compared to existing WPT receivers, can be expected.

ACKNOWLEDGEMENT

This work was supported by the Hong Kong Research Grant Council under the GRF project 17204318